\documentclass[12pt,a4paper]{elsart}
\usepackage{ifthen,graphics}
\usepackage{color}
\usepackage{cite}
\usepackage{epsfig}

\definecolor{dgreen}{rgb}{.0,.7,.0}

\newcommand{\Fig}[1]{Fig.~\ref{#1}}
\newcommand{\Ref}[1]{Ref.~\cite{#1}}

\newcommand{\Tab}[1]{Tab.~\ref{#1}}

\newcommand{\ie}{{\em i.e.}}

\newcommand{\ith}{\ensuremath{i^\mathrm{th}}}

\newcommand{\VA}[3]{\ifthenelse{\equal{#2}{#3}}
{\ensuremath{#1\pm#2}}{\ensuremath{#1\,^{+#2}_{-#3}}}}

\newcommand{\pvec}{\ensuremath{\mathbf{p}}}
\newcommand{\rxy}{\ensuremath{r_{xy}}}

\newcommand{\xvec}{\ensuremath{\mathbf{x}}}

\newcommand{\DKeIII}{\ensuremath{K_{e3}}}

\newcommand{\DKmuIII}{\ensuremath{K_{\mu 3}}}

\def \gev  {{\rm \,Ge\kern-0.125em V}}
\def \mev  {{\rm \,Me\kern-0.125em V}}
\def \kev  {{\rm \,ke\kern-0.125em V}}
\def \ev   {{\rm \,e\kern-0.125em V}}

\newcommand{\dTCA}{\ensuremath{d_{\mathrm{TC}}}}
\newcommand{\dtTCA}{\ensuremath{d_{\perp,\,\mathrm{TC}}}}

\newcommand{\xc}{\ensuremath{\mathbf{x}_\mathrm{c}}}
\newcommand{\pc}{\ensuremath{\mathbf{p}_\mathrm{c}}}
\newcommand{\lc}{\ensuremath{l_\mathrm{c}}}
\newcommand{\dc}{\ensuremath{d_\mathrm{c}}}

\newcommand{\klpln}{\mbox{$K_{L}\to\pi^{\pm}\ell^{\mp}\nu$}}
\newcommand{\klpmn}{\mbox{$K_{L}\to\pi\mu\nu$}}
\newcommand{\semil}{\mbox{$K_{\mu3}$}}
\renewcommand{\vec}[1]{\mbox{\boldmath{$\rm#1$}}}

\newcommand{\fphat}{\mbox{$\tilde{f}_+(t)$}}

\newcommand{\fzhat}{\mbox{$\tilde{f}_0(t)$}}

\newcommand{\miss}{\mbox{$E_{miss}-p_{miss}$}}

\newcommand{\kspp}{\mbox{$K_{S}\to\pi^+\pi^- $}}

\newcommand{\klpen}{\mbox{$K_{L}\to\pi^{\pm}e^{\mp}\nu$}}
\newcommand{\aff}[2]{Dipartimento di Fisica dell'Universit\`a #1 e Sezione INFN, #2, Italy.}
\newcommand{\affd}[1]{Dipartimento di Fisica dell'Universit\`a e Sezione INFN, #1, Italy.}

\renewcommand{\thesection}{\arabic{section}}%
\makeatletter
\renewcommand{\section}{\@startsection{section}%
{1}%
{0mm}%
{0.95\baselineskip}
{0.5\baselineskip}
{\normalfont\large\bf\mathversion{bold}}}%
\makeatother
\makeatletter
\renewcommand{\subsection}{\@startsection{subsection}%
{2}%
{0mm}%
{0.95\baselineskip}
{0.5\baselineskip}
{\normalfont\normalsize\bf\mathversion{bold}}}%
\makeatother

\def\ifm#1{\relax\ifmmode#1\else$#1$\fi}  

\def\x{\ifm{\times}}  
\def\pt#1,#2,{\ifm{#1\x10^{#2}}}
\def\up#1{\ifm{^{#1}}}  

\def\ab{\ifm{\sim}}

\def\to{\ifm{\rightarrow}} 
\def\kl{\ifm{K_L}}   
\def\ks{\ifm{K_S}}

\def\po{\ifm{\pi^0}}
\def\pic{\ifm{\pi^+\pi^-}}  
   \def\ff{$\phi$--factory}
\def\rmk{\rm\kern.5mm }   \def\dif{\hbox{d}}
  \def\minus{$-$}

\def\bye{\end{document}}

\newcommand{\lamp}{\mbox{$\lambda'_+$}}
\newcommand{\lampp}{\mbox{$\lambda''_+$}}
\newcommand{\lamo}{\mbox{$\lambda_0$}}
\newcommand{\lamop}{\mbox{$\lambda'_0$}}
\newcommand{\lamopp}{\mbox{$\lambda''_0$}}
\newcommand{\jth}{\ensuremath{j^\mathrm{th}}}
\def\dlp{\hbox{$\delta\lambda_+'$}}
\def\dlpp{\hbox{$\delta\lambda_+''$}} \def\dop{\hbox{$\delta\lambda_0'$}}
\def\dopp{\hbox{$\delta\lambda_0''$}}

\begin{document}
\begin{frontmatter}
\title{\mathversion{bold}Measurements of the form-factors slopes of 
       \kl\ \to\ $\pi\mu\nu$ decay with the KLOE Detector}
\collab{The KLOE Collaboration}
\author[Na]{F.~Ambrosino},
\author[Frascati]{A.~Antonelli},
\author[Frascati]{M.~Antonelli},
\author[Frascati]{F.~Archilli},
\author[Roma3]{C.~Bacci},
\author[Karlsruhe]{P.~Beltrame},
\author[Frascati]{G.~Bencivenni},
\author[Frascati]{S.~Bertolucci},
\author[Roma1]{C.~Bini},
\author[Frascati]{C.~Bloise},
\author[Roma3]{S.~Bocchetta},
\author[Roma1]{V.~Bocci},
\author[Frascati]{F.~Bossi},
\author[Roma3]{P.~Branchini},
\author[Roma1]{R.~Caloi},
\author[Frascati]{P.~Campana},
\author[Frascati]{G.~Capon},
\author[Na]{T.~Capussela},
\author[Roma3]{F.~Ceradini},
\author[Frascati]{S.~Chi},
\author[Na]{G.~Chiefari},
\author[Frascati]{P.~Ciambrone},
\author[Frascati]{E.~De~Lucia},
\author[Roma1]{A.~De~Santis},
\author[Frascati]{P.~De~Simone},
\author[Roma1]{G.~De~Zorzi},
\author[Karlsruhe]{A.~Denig},
\author[Roma1]{A.~Di~Domenico},
\author[Na]{C.~Di~Donato},
\author[Pisa]{S.~Di~Falco},
\author[Roma3]{B.~Di~Micco},
\author[Na]{A.~Doria},
\author[Frascati]{M.~Dreucci\corauthref{cor1}},
\author[Frascati]{G.~Felici},
\author[Frascati]{A.~Ferrari},
\author[Frascati]{M.~L.~Ferrer},
\author[Frascati]{G.~Finocchiaro},
\author[Roma1]{S.~Fiore},
\author[Frascati]{C.~Forti},
\author[Roma1]{P.~Franzini},
\author[Frascati]{C.~Gatti},
\author[Roma1]{P.~Gauzzi},
\author[Frascati]{S.~Giovannella},
\author[Lecce]{E.~Gorini},
\author[Roma3]{E.~Graziani},
\author[Pisa]{M.~Incagli},
\author[Karlsruhe]{W.~Kluge},
\author[Moscow]{V.~Kulikov},
\author[Roma1]{F.~Lacava},
\author[Frascati]{G.~Lanfranchi},
\author[Frascati,StonyBrook]{J.~Lee-Franzini},
\author[Karlsruhe]{D.~Leone},
\author[Frascati]{M.~Martini},
\author[Na]{P.~Massarotti},
\author[Frascati]{W.~Mei},
\author[Na]{S.~Meola},
\author[Frascati]{S.~Miscetti},
\author[Frascati]{M.~Moulson},
\author[Frascati]{S.~M\"uller},
\author[Frascati]{F.~Murtas},
\author[Na]{M.~Napolitano},
\author[Roma3]{F.~Nguyen},
\author[Frascati]{M.~Palutan},
\author[Roma1]{E.~Pasqualucci},
\author[Roma3]{A.~Passeri},
\author[Frascati,Energ]{V.~Patera},
\author[Na]{F.~Perfetto},
\author[Lecce]{M.~Primavera},
\author[Frascati]{P.~Santangelo},
\author[Na]{G.~Saracino},
\author[Frascati]{B.~Sciascia},
\author[Frascati,Energ]{A.~Sciubba},
\author[Pisa]{F.~Scuri},
\author[Frascati]{I.~Sfiligoi},
\author[Frascati,Novo]{A.~Sibidanov},
\author[Frascati]{T.~Spadaro},
\author[Roma1]{M.~Testa},
\author[Roma3]{L.~Tortora},
\author[Roma1]{P.~Valente},
\author[Karlsruhe]{B.~Valeriani},
\author[Frascati]{G.~Venanzoni},
\author[Frascati]{R.Versaci},
\author[Frascati,Beijing]{G.~Xu}

\address[Virginia]{Physics Department, University of Virginia, Charlottesville, VA, USA.}
\address[Frascati]{Laboratori Nazionali di Frascati dell'INFN, Frascati, Italy.}
\address[Karlsruhe]{Institut f\"ur Experimentelle Kernphysik, Universit\"at Karlsruhe, Germany.}
\address[Lecce]{\affd{Lecce}}
\address[Na]{Dipartimento di Scienze Fisiche dell'Universit\`a ``Federico II'' e Sezione INFN, Napoli, Italy}
\address[Energ]{Dipartimento di Energetica dell'Universit\`a ``La Sapienza'', Roma, Italy.}
\address[Roma1]{\aff{``La Sapienza''}{Roma}}
\address[Roma2]{\aff{``Tor Vergata''}{Roma}}
\address[Roma3]{\aff{``Roma Tre''}{Roma}}
\address[Moscow]{Institute for Theoretical and Experimental Physics, Moscow, Russia.}
\address[Pisa]{\affd{Pisa}}
\address[StonyBrook]{Physics Department, State University of New York at Stony Brook, NY, USA.}
\address[Novo]{Permanent address: Budker Institute of Nuclear Physics, Novosibirsk, Russia.}
\address[Beijing]{Permanent address: Institute of High Energy Physics, CAS, Beijing, China.}

\begin{flushleft}
\corauth[cor1]{cor1}{\small $^1$ Corresponding author: Mario Antonelli
INFN - LNF, Casella postale 13, 00044 Frascati (Roma), 
Italy; tel. +39-06-94032728, e-mail Mario.Antonelli@lnf.infn.it}
\end{flushleft}
\begin{flushleft}
\corauth[cor2]{cor2}{\small $^2$ Corresponding author: Marianna Testa
Dipartimento di Fisica dell'Universit\`a ``La Sapienza'' e Sezione INFN, Roma, Italy.
Italy; tel. +39-06-94032696, e-mail Marianna.Testa@lnf.infn.it}
\end{flushleft}

\begin{abstract}
 We present a measurement of the $K$-$\pi$  form-factor parameters for the decay 
 \klpmn. We use 328 pb$^{-1}$ of data collected in 2001 and 2002, corresponding to
 $\sim $ 1.8 million \semil\ events.  
 Measurements of semileptonic form factors provide information about the dynamics of the
 strong interaction and are necessary for evaluation 
 of the phase-space integral $I^\mu_K$ needed to measure the CKM matrix element $|V_{us}|$ 
 for \klpmn\ decays and to test lepton universality in kaon decays. 
 Using a new parameterization for the vector and scalar form factors we find
 $\lambda_+$=\pt(25.6\pm 0.4_{\rm{stat.}}\pm 0.3_{\rm{syst.}}),-3, and
 $\lambda_0$=\pt(14.3\pm 1.7_{\rm{stat.}}\pm 1.1_{\rm{stat.}}),-3,. In the more usual quadratic expansion of the form factor the above result is corresponds to $\lambda'_+=\lambda_+$, $\lambda''_+=2\lambda^2_+$, $\lambda'_0=\lambda_0$ and $\lambda''_0=(\lambda^2_0+0.000416)/2$. Our results, together with recent lattice calculations of $f_\pi$, $f_K$ and $f(0)$, satisfy the Callan-Trieman relation.

\end{abstract}
\end{frontmatter}

\section{Introduction}
\begin{figure}[ht]
  \label{fig:diag}  
  \begin{center}
    \resizebox{0.3\textwidth}{!}{\includegraphics{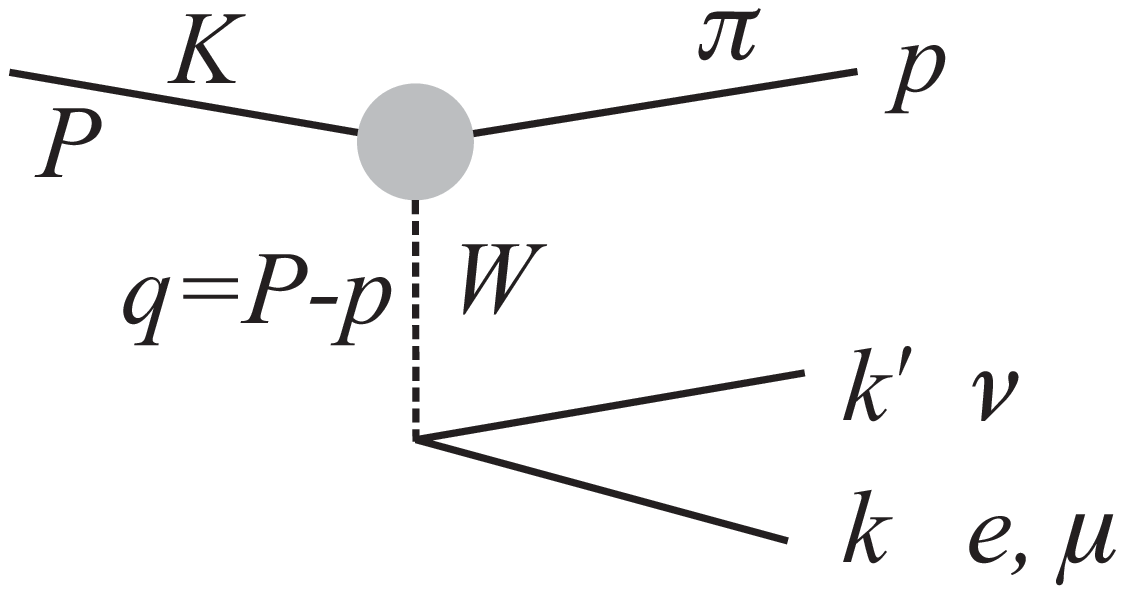}}
  \end{center}
  \caption{Amplitude for \klpln. The gray region indicates the $K\to\pi W$ vertex structure. }
\end{figure}

\parskip 2mm
%
 Semileptonic kaon decays, \klpln, (\Fig{fig:diag}) offer possibly the cleanest way to obtain an 
 accurate value of the Cabibbo angle, or better, $V_{us}$. 
 Since $K\to \pi$ is a $0^-\to 0^-$  transition, only the vector part of the hadronic weak current
 has a non vanishing contribution. 
 Vector transitions are protected by the Ademollo-Gatto theorem against SU(3) breaking corrections to lowest order in $m_s$ (or $m_s-m_{u,d}$). At present, the largest uncertainty in calculating $V_{us}$ from
 the decay rate, is due to the difficulties in  computing the matrix element
 $\langle\pi|J^{\rm had}_{\alpha}|K\rangle$.
Using the notation of \Fig{fig:diag}, Lorentz invariance requires that the above matrix element have the form 
\begin{equation}
\label{hc}
 <\pi|J^{\rm had}_\alpha|K>=\left((P+p)_\alpha f_+(t)+(P-p)_\alpha f_-(t)\right)
\end{equation}
where $t=(P-p)^2=(k+k')^2=M^2+m^2-2ME_\pi$ is the only $L$-invariant variable.
The form factors, FF, $f_+(t)$ and $f_-(t)$ account for the non point like structure of the hadrons and the values of the FF's at $t=0$ differs from one because of $SU(3)$ corrections, \ie\ because pions and kaons have different structure.
The $P-p=k+k'$ term in the $f_-$ form factor, acting on the lepton term gives the lepton mass and is therefore negligible for $K_{\rm e3}$ decays. The $f_-$ form factor must be 
retained for $K_{\mu3}$ decays. It has become costumary to introduce a scalar form 
factor $f_0(t)$ according to 
$$\langle\pi(p)|\bar u\gamma_\alpha s|K(P)\rangle=f(0)\left((P+p)_\alpha\,\tilde f_+(t)+(P-p)_\alpha\left(\tilde f_0(t){\Delta_{K\pi}\over t}-\tilde f_+(t){\Delta_{K\pi}\over t}\right)\right),$$
with $\Delta_{K\pi}=M^2-m^2$. The $f_+$ and $f_0$ FFs must have the same value at $t=0$. We have therefore factored out a term $f(0)$. The functions $\tilde f_+(t)$ and $\tilde f_0(t)$ are both unity at $t=0$.
If the FF are expanded in powers of $t$ up to $t^2$ as
\begin{equation}
\tilde f_{+,0}(t) = 1 + \lambda'_{+,0}~\frac{t}{m^2}+\frac{1}{2}\;\lambda''_{+,0}\,\left(\frac{t}{m^2}\right)^2
  \label{eq:ff2}
\end{equation}
four parameters: $\lambda'_+,\ \lambda''_+,\ \lambda'_0$ and $\lambda''_0$ need to be determined from the decay spectrum in order to be able to compute the phase space integral which appears in the formula for the partial decay width. The problems with the four parameters above is the large correlations, in particular \minus99.96\% between $\lambda'_0$ and $\lambda''_0$ and \minus97.6\% between $\lambda'_+$ and $\lambda''_+$. It is not therefore possible to obtain meaningful results for the scalar FF parameters, see the appendix, eq. \ref{4corr}.

It is experimentally well established in ${\kl}_{e3}$ decays \cite{KTeV:ff,NA48:ff,KLOE:ff}, that the vector form factor is equally described by a pole form:
\begin{equation}
\fphat ={M_V^2\over M_V^2-t}.
  \label{eq:pole}
\end{equation}
which expands to $1+t/M_V^2+(t/M_V^2)^2$, neglecting power of $t$ greater than 2.

Recent results on $K_{e3}$ show that the vector form factor is dominated by the closest vector $(q\bar q)$ state with one strange and one light quark (or $K$-$\pi$ resonance in an older language) and are in good agreement with the results from fitting with a vector from factor $\tilde f_+(t)$ as in eq. \ref{eq:ff2} \cite{pf}. There is however better consistency between the pole than the quadratic expansion fits, due mostly by the additional fluctuation introduced by the correlation, \minus95\% between \lamp\ and \lampp. The results are also consistent with predictions from a dispersive approach \cite{sternV,pich1}. We will therefore mostly use the following form for the vector form factor:
\begin{equation}
\fphat =1+\lambda_+{t\over m^2}+\lambda_+^2\left(t\over m^2\right)^2.
  \label{fplus}
\end{equation}
$K_{\mu3}$ decay pion spectrum measurements, reported in \cite{KTeV:ff,NA48:ff2,istra2}, have no sensitivity to $\lamopp$, see apendix. Therefore, all authors have fitted for a linear scalar form factor:
\begin{equation}
\fzhat =1+\lambda_0 {t\over m^2}.
  \label{fzero}
\end{equation}
Because of correlation this leads to incorrect answers for the value of \lamop\ which comes out of the fit increased by \ab3.5\x the coefficient of the $t^2$ term. To clarify this situation it is necessary to obtain a form for \fzhat\ with $t$ and $t^2$ terms but with only one parameter.
 The Callan-Treiman relation\cite{ct} fixes the
 value of scalar form factor at $t=\Delta_{K\pi}$ (the so called Callan-Trieman point) to the ratio of the pseudoscalar 
 decay constants $f_K/f_\pi$. This relation is slightly modified by SU(2) breaking
 corrections\cite{dct}:
\begin{equation}
\tilde{f}_0(\Delta_{K\pi})=\frac{f_K}{f_\pi}\:{1\over f(0)}+\Delta_{CT},\quad \Delta_{CT}\simeq\pt-3.4,-3,
  \label{eq:ct}
\end{equation}
 A recent parametrization for the scalar form factor \cite{stern}
 allow to take into account the constraint given by the Callan-Treiman relation:
\begin{equation}
\tilde{f}_0(t)=\exp\left(\frac{t}{\Delta_{K\pi}} \log C-G(t)\right)
  \label{eq:stern}
\end{equation}
 where $G(t)$ is obtained using a  dispersion relation subtracted at $t=\Delta_{K\pi}$,
 such that $C=\tilde{f}_0(\Delta_{K\pi})$.
As suggested in \cite{stern}, a good approximation to eq. \ref{eq:stern} is
\begin{equation}
\fzhat =1+\lambda_0 {t\over m^2}+{\lambda_0^2+0.000416\over 2}\left({t\over m^2}\right)^2.
  \label{fzeroa}
\end{equation}
This result is quite similar to ref. \cite{pich2}. \noindent
With KLOE, we can measure the pion energy spectrum ($t=M^2+m^2-2ME\pi$) spectrum since the value of \kl\ momentum is known at a \ff. $\pi-\mu$ separation is however very difficult at low energy. Attempts to distinguish pions and muons result in a loss of events of more than a factor of 2 and introduce severe systematic uncertainties. Therefore we use the neutrino spectrum that can be obtained without $\pi-\mu$ identification.

\section{The KLOE detector}
\label{sec:kloedet}
The KLOE detector consists of a large cylindrical 
drift chamber (DC), surrounded by a 
lead scintillating-fiber electromagnetic calorimeter 
(EMC). A superconducting coil around the calorimeter 
provides a 0.52 T field. 
The drift chamber \cite{KLOE:DC} is
4~m in diameter and 3.3~m long.
The momentum 
resolution is $\sigma_{p_{\perp}}/p_{\perp}\approx 0.4\%$. 
Two-track vertices are reconstructed with a spatial 
resolution of $\sim$ 3~mm. 
The calorimeter \cite{KLOE:EmC} 
is divided into a barrel and two endcaps. It covers 98\% of the solid 
angle. Cells close in time and space are grouped into 
calorimeter clusters. The energy and time resolutions 
are $\sigma_E/E = 5.7\%/\sqrt{E\ {\rm(GeV)}}$ and  
$\sigma_T = 57\ {\rm ps}/\sqrt{E\ {\rm(GeV)}}\oplus100\ {\rm ps}$, 
respectively.
The KLOE trigger \cite{KLOE:trig} 
uses calorimeter and chamber information. For this 
analysis, only the calorimeter signals are used. 
Two energy deposits above threshold ($E>50$ MeV for the 
barrel and  $E>150$ MeV for the endcaps) are required. 
Recognition and rejection of cosmic-ray events is also 
performed at the trigger level. Events with two energy 
deposits above a 30 MeV threshold in the outermost 
calorimeter plane are rejected.

\section{Analysis}
\label{sec:analysis}
Candidate  \kl\ events are tagged by the presence of a \kspp\ decay. 
The \kl\ tagging algorithm is fully described in Refs.~\cite{KLOE:brlnote} and~\cite{KLOE:brl}.
The \kl\ momentum, $p_{\kl}$, is obtained from the kinematics of the 
$\phi\to\ks\kl$ decay, using the reconstructed \ks\ direction and the known value of $\pvec_{\phi}$. 
The resolution  is dominated by the beam-energy spread, and amounts to about 
0.8 MeV/c. The position of the $\phi$\ production point, $\xvec_\phi$, is determined 
as the point of closest approach of the \ks\ path to the beam line. The \kl\ line of flight 
({\em tagging line}) is given by the \kl\ momentum,  
$\pvec_{\kl} = \pvec_\phi-\pvec_{\ks}$ and the production vertex position, 
$\xvec_\phi$.
 All relevant tracks in the chamber, after removal of those from the \ks\ 
decay and their descendants, are extrapolated to their points of closest
approach to the {\em tagging line}.

For each track candidate, we evaluate the point of closest approach to the 
{\em tagging line}, \xc, and the distance of closest approach, \dc. The
momentum \pc\ of the track at \xc\ and the extrapolation length, \lc, are also
computed. Tracks satisfying $\dc< a \rxy + b$, with $a=0.03$ and $b=3$ cm, and 
$-20<\lc<25$ cm are accepted as \kl\ decay products. \rxy\ the distance of the vertex from the beam line. For each charge sign we chose the track with the smallest value of \dc\ as a \kl\ decay product and from them we reconstruct the decay vertex. 
The combined tracking and vertexing efficiency for \semil\ is about 54\%.
This above value is determined from data as described in \Ref{KLOE:brlnote}.
Events are retained if the vertex is in the fiducial volume
$35<\rxy<150$ cm and $|z|<120$ cm.

Background from \kl\to\pic,\ \pic\po\ is easily removed by loose kinematic cuts.
The largest background is due to \klpen\ decays, possibly followed by early $\pi\to\mu e$ decay in flight. For all candidate $K_{\mu3}$ events we compute $\Delta_{\pi e}$, the lesser between $|\miss |$ assuming the decay particles are $\pi e$ or $\mu e$.
We retain events only if this variable is greater than 10 \mev.
After the above kinematic cuts the efficiency for the signal is about 96\% and the purity is about 80\%. 

\begin{figure}[!htb]
  \label{fig:ellipse}  
  \begin{center}
    \resizebox{0.48\textwidth}{!}{\includegraphics{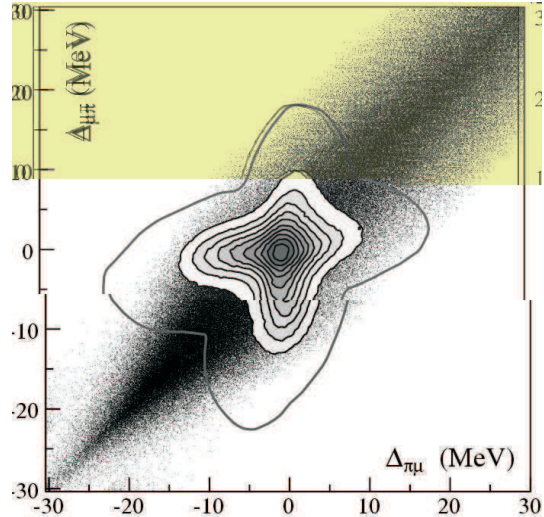}}
  \end{center}
  \caption{ $E_{\rm miss}(\pi^{+},\mu^{-})$ versus $E_{\rm miss}(\pi^{-},\mu^{+})$ distribution from Monte Carlo.
  \kl\ \to $\pi\mu\nu$ (gray scale) and background (black dot).}
\end{figure}

A further cut  on the distribution $E_{\rm miss}(\pi^{+},\mu^{-})$-$E_{\rm miss}(\pi^{-},\mu^{+})$ shown in  Fig.~\ref{fig:ellipse}  for \kl\ \to\ $\pi \mu \nu$ and background events respectively, is applied
After the kinematic cuts described above, the contamination, dominated by \kl\to $\pi e\nu$ decay is $\sim$4\%. Particle identification (PID) based on 
calorimeter information further reduces the contamination by $\sim$2.

\begin{figure}[!htb]
  \label{fig:tofcorr}  
  \begin{center}
    \resizebox{0.48\textwidth}{!}{\includegraphics{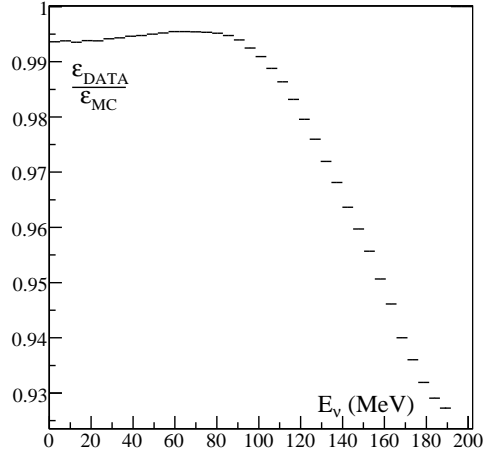}}
  \end{center}
  \caption{  Cluster efficiency correction versus $E_\nu$.}
\end{figure}

Tracks are required to be associated with EMC clusters. 
We define two  variables: \dTCA, the distance from the extrapolated track entry point in the calorimeter to the cluster centroid and \dtTCA, the component of this distance in the 
plane orthogonal to the momentum of the track at the entry position.
We accept tracks with \dtTCA\ $<$ 30~cm.
The cluster efficiency is obtained from the KLOE Monte Carlo (MC), corrected 
with the ratio of data and MC efficiencies obtained from control samples.
These samples, of 86\% and 99.5\% purity, are obtained from \semil\ and $K_{e3}$ 
selected by means of kinematics and independent calorimeter information.
The cluster efficiency correction versus $E_\nu$ is shown in Fig.~\ref{fig:tofcorr}.

For each \kl\ decay track with an associated cluster we define the variable: $\Delta t_i = t_{\rm cl}-t_{i} ,\ (i=\pi,\ e)$ in which $t_{\rm cl}$ is the cluster time and $t_{i}$ is the expected time of flight, 
evaluated according to a well defined mass hypothesis. We evaluate $t_{i}$ including also the time from 
the entry point to the cluster centroid~\cite{KLOE:kssemi}.
We determine the $e^+e^-$ collision time, $t_0$, using the clusters from the \ks.

For this pupose, for each \kl\ decay track with an associated cluster, we define the variable: 
$\Delta t_i = t_{\rm cl}-t_{i} ,~(i = \pi,~ e) $ in which $t_{\rm cl}$ is the cluster time 
and $t_{i}$ is the expected time of flight, evaluated according to a well-defined mass 
hypothesis. The evaluating of $t_{i}$ includes the propagation time from 
the entry point to the cluster centroid~\cite{KLOE:kssemi}.

An effective way to select the correct mass assignment, $\pi e$ or $e\pi$,
is obtained by choosing the lesser of $|\Delta t_{\pi^+}-\Delta t_{e^-}|$ and
$|\Delta t_{\pi^-}-\Delta t_{e^+}|$.
After the mass assignment has been made, we consider the variable
 $$R_{\rm TOF}=\left(\frac{\displaystyle \Delta^\pi +\Delta^e}{\displaystyle 2\sigma_1}\right)^2+\left(\frac{\displaystyle \Delta^\pi -\Delta^e }{\displaystyle 2\sigma_2} \right).$$

\begin{figure}[!htb]
  \label{fig:nntof}  
  \begin{center}
    \resizebox{0.48\textwidth}{!}{\includegraphics{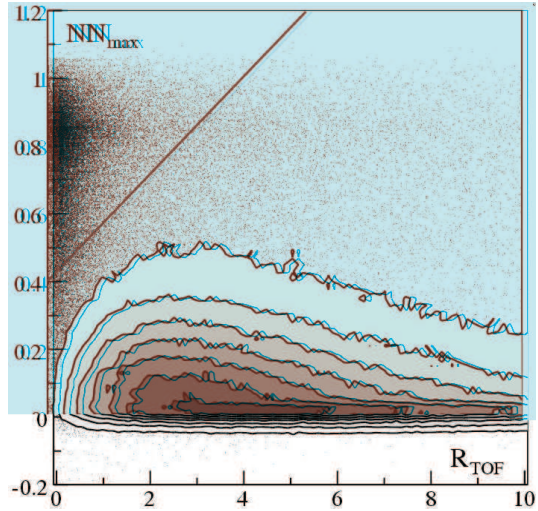}}
  \end{center}
  \caption{ $\rm{NN_{max}}$ vs $R_{\rm TOF}$ distribution (see text for the definitions)
   from Monte Carlo.
  \kl\ \to $\pi\mu\nu$ (gray scale) and background (black dot).}
\end{figure}

\begin{figure}[!htb]
  \label{fig:purity}  
  \begin{center}
    \resizebox{0.48\textwidth}{!}{\includegraphics{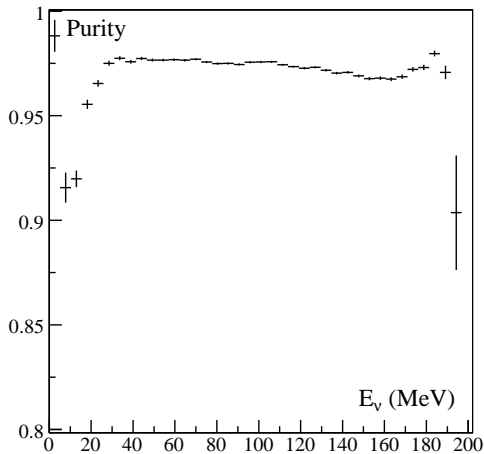}}
  \end{center}
  \caption{ Purity versus $E_\nu$.}
\end{figure}

 Additional informations are provided by the energy deposition
 in the calorimeter and the cluster centroid depth. These measurements have been input
to a Neural Network (NN).
 The value of $R_{\rm TOF}$ and that of
  the maximum of the NN outputs ($\rm{NN_{max}}$) for the two charge hypotehsis are 
 shown in Fig.~\ref{fig:nntof} for data and MC. We retain events with
$R_{\rm TOF}<1/6\rm{NN_{max}}+0.4$ as indicated in Fig.~\ref{fig:nntof}.
 The resulting  purity of the sample is $\sim \,97.5\%$, almost uniform in the fit range $21 MeV<E_\nu<166 MeV$ (Fig.~\ref{fig:purity}).

The form-factor parameters are obtained by fitting the $E_{\nu}$ distribution of the selected
 events in the range $21<E_\nu<166$ MeV.
After subtracting the residual background as estimated from MC, we perform the fit
using the following formula:
\begin{equation}
  \label{fitfunc}
  \frac{dN}{dt}(i)=N_0\sum_{j=1}^{20} A(i,j)\times~\rho(j,\lamp,\lampp, \lamo )\times~\epsilon_{tot}(j)
  \times~F_{FSR}(j)
\end{equation}
where $\rho(j,\lamp,\lampp,\lamo)$ is the three-body differential decay width,
 and $A(i,j)$ is the probability that an event with true value of $E_\nu$ 
in the \jth\ bin has a reconstructed value in the \ith\ bin.
The chosen bin size is
 5.18 MeV, which corresponds to about 
1.7 $\sigma_{E \nu}$, where $\sigma_{E_\nu}$ is the resolution on the neutrino energy.

$F_{FSR}$ is the correction due to final state radiation.
It is evaluated using the KLOE  MC  simulation, GEANFI ~\cite{KLOE:offline},
where FSR processes are simulated according the procedures described in \Ref{KLOE:gattip}.
FSR affects $t$-distribution mainly for high energy pions,
\ie\ for low $t$, where the correction is 3-5\%.   
The slopes  \lamp, \lampp\ and \lamo are free parameters in the fit while the $N_0$
constant is the total number of signal events.

\section{Systematic uncertainties}
The systematic errors due to the evaluation of corrections, data-MC inconsistencies, 
result stability, momentum mis-calibration, and background contamination are summarized 
in \Tab{tabsyst}.

\begin{table}[hbt]
  \begin{center}
    \begin{tabular}{|c|c|c|c|}
      \hline 
  Source &  $\delta\lamp\times 10^{3}$ & $\delta\lampp\times 10^{3}$ & $\delta\lamo\times 10^{3}$ \\
      \hline
      Tracking     &    1.60   &  0.47  & 0.86   \\  
      Clustering   &    1.50   &  0.32  & 1.45   \\
      TOF + NN     &    2.23   &  1.16  & 1.45   \\
      p-scale &    1.10   &  0.71  & 0.81   \\
      p-resolution &    0.61   &  0.21  & 0.01   \\
      \hline
      Total        &    3.37   &  1.50  & 2.37   \\
      \hline
    \end{tabular}  
    \caption{Summary of systematic uncertainties on \lamp, \lampp, \lamo.}
    \label{tabsyst}    
  \end{center}
\end{table}
%

We evaluate the systematic uncertainties on the tracking efficiency corrections by checking 
stability of the result when the selection of tracks is modified.
We proved the validity of the method by comparing the efficiencies from data 
and MC control samples, and from the MC {\em truth}~\cite{KLOE:kssemi}.
The uncertainty on the tracking efficiency correction is dominated by sample
statistics and by the variation of the results observed using different criteria to
identify tracks from \kl\ decays. Its statistical error is taken into account in
the fit.
 We study the effect of differences in the resolution with which the variable
 \dc\ is reconstructed in data and in MC, and the possible bias introduced in the 
 selection of the control sample, by varying the values of the 
 cuts made on this variable when associating tracks to \kl\ vertexes.
 For each variation, corresponding to a maximal change of the tracking efficiency
 of about $\pm$10\%, we evaluate the complete tracking-efficiency correction 
 and measure the slope parameters.
We observe  changes of 1.6$\times10^{-3}$, 
  0.47$\times10^{-3}$, and 0.86$\times10^{-3}$ 
for  \lamp\, \lampp\ and \lamo\ , respectively.
We find a smaller uncertainty by comparing the efficiencies 
from data and MC control sample, and MC {\em truth}. 
However, we assume conservatively the changes in the result observed by
varying the cut on \dc\ as a systematic uncertainty.

 As for tracking, we evaluate the systematic uncertainties on the clustering efficiency
 corrections by checking stability of the result when the track-to-cluster association
 criteria are modified.
 Also in this case the uncertainty on the clustering efficiency corrections is dominated
 by sample 
 statistics and by the variation of the results observed using different criteria for the 
 track-to-cluster association.  We take into account its statistical error in the fit.
 The most effective variable in the definition of track-to-cluster association
 is the transverse distance, \dtTCA. We vary the cut on \dtTCA\ in a wide range from
 15 cm to 100 cm,
 corresponding to a change in efficiency of about 19\%.
For each configuration, we obtain the complete track extrapolation and clustering 
efficiency correction and we use it to evaluate the slopes.
We observe a corresponding  changes of  1.5$\times10^{-3}$, 0.32$\times10^{-3}$ and 1.45$\times10^{-3}$
for \lamp\,\lampp\  and \lamo, respectively.

 We study the uncertainties  on the efficiency of the and on the background
 evaluation by repeating the measurement on samples with modified PID and
 kinematic cut values, corresponding to a variation of the cut efficency from 
  90\% to 95\%. This allows to vary the background contamination
 from 1.5\% to 4.5\%.    
We observe a corresponding  changes of  2.23$\times10^{-3}$, 1.16$\times10^{-3}$ and 1.45$\times10^{-3}$
for \lamp\,\lampp\  and \lamo, respectively.

The effect of the momentum scale and the momentum resolution have also been
 considered.
We conservatively assume a momentum scale uncertainty of 0.1\%,
well above the known KLOE scale accuracy.
We observe a corresponding  changes of  1.1$\times10^{-3}$, 0.71$\times10^{-3}$ and 0.81$\times10^{-3}$
for \lamp\,\lampp\  and \lamo, respectively.

We investigate the effect of momentum resolution by changing the value
 of the $E_{\nu}$   resolution by 3\% as studied in \cite{KLOE:ff}.
 The corresponding absolute changes are 0.61$\times10^{-3}$,
0.21$\times10^{-3}$ and 0.01$\times10^{-3}$  for \lamp\,\lampp\ amd \lamo\ respectively.

\section{Results and interpretation}
\begin{figure}[!htb]
  \label{fig:fit_res}  
  \begin{center}
    \resizebox{0.7\textwidth}{!}{\includegraphics{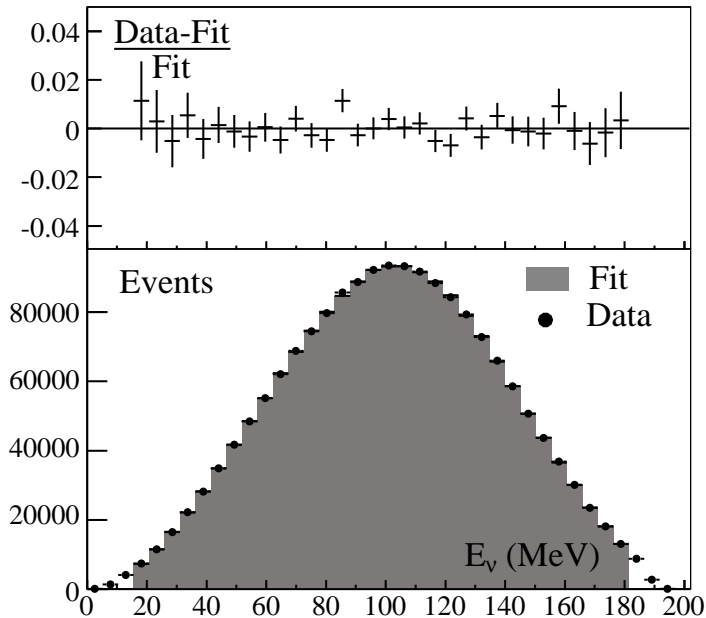}}
  \end{center}
  \caption{ Residuals of the fit (top plot) and $E_nu$ distribution for data events superimposed on the fit result (bottom plot)}
\end{figure}

 About 1.7 Milion of \DKmuIII\ events have been selected.
 We fit data, using the quadratic parametrization for the
 vector form factor and linear parametrization for the scalar form factor.
The results are shown in Fig.~\ref{fig:fit_res}.
 We obtain:\\
\parbox{.5\textwidth}{\begin{eqnarray*}
  \lamp &=& (22.3\pm 9.8_{\rm{stat.}}\pm 3.4_{\rm{syst.}}) \times \rm{10^{-3}} \\
  \lampp &=& (4.8\pm 4.9_{\rm{stat.}}\pm 1.5_{\rm{syst.}}) \times \rm{10^{-3}} \\
  \lambda_0 &=& (9.1\pm 5.9_{\rm{stat.}}\pm 2.4_{\rm{stat.}}) \times \rm{10^{-3}} 
\end{eqnarray*}}\kern1cm
\parbox{.35\textwidth}{\begin{displaymath}
\left(
\begin{array}{ccc}
 1 & -0.97 & 0.81 \\
   &     1 & -0.91 \\
   &       & 1     \\
\end{array}
\right)
\end{displaymath}}\\
with $\chi^2/{\rm dof}=19/29$. The  correlations are given by the matrix.

Improved accuracy is obtained combining the above results
 with those from the \DKeIII\ analysis\cite{KLOE:ff}:
\begin{eqnarray*}
  \lamp &=& (25.5\pm 1.5_{\rm{stat.}}\pm 1.0_{\rm{syst.}}) \times \rm{10^{-3}} \\
  \lampp &=& (1.4\pm 0.7_{\rm{stat.}}\pm 0.4_{\rm{syst.}}) \times \rm{10^{-3}} \\
\end{eqnarray*}
We then find:\\
\parbox{.5\textwidth}{\begin{eqnarray*}
  \lamp &=& (25.6\pm 1.5_{\rm{stat.}}\pm 0.9_{\rm{syst.}}) \times \rm{10^{-3}} \\
  \lampp &=& (1.5\pm 0.7_{\rm{stat.}}\pm 0.4_{\rm{syst.}}) \times \rm{10^{-3}} \\
  \lambda_0 &=& (15.4\pm 1.8_{\rm{stat.}}\pm 1.1_{\rm{stat.}}) \times \rm{10^{-3}} 
\end{eqnarray*}}\kern1cm
\parbox{.35\textwidth}{\begin{displaymath}
\left(
\begin{array}{ccc}
 1 & -0.95 & 0.29 \\
   &     1 & -0.38 \\
   &       & 1     \\
\end{array}
\right)
\end{displaymath}}\\
with $\chi^2/{\rm dof}=2.3/2$ with the correlation given in the matrix on the right.\\
Finally, we fit data using the parameterization for the scalar form factor given in eq. \ref{fzeroa}. Improved accuracy is obtained by using the pole parameterization for the 
vector form factor, truncated as in eq. \ref{fplus}. Dropping the ``$\;'\;$'' indexes, we find\\
\parbox{.5\textwidth}{\begin{eqnarray*}
  \lambda_+ &=& (25.6\pm 0.4_{\rm{stat.}}\pm 0.3_{\rm{syst.}}) \times \rm{10^{-3}} \\
  \lambda_0 &=& (14.3\pm 1.7_{\rm{stat.}}\pm 1.1_{\rm{stat.}}) \times \rm{10^{-3}} 
\end{eqnarray*}}\kern1cm\parbox{.35\textwidth}{\begin{displaymath}
\left(
\begin{array}{cc}
 1 & -0.26  \\
   &     1  \\
\end{array}
\right)
\end{displaymath}}\\
with $\chi^2/{\rm dof}=2.56/3$ and the correlation given at right. We remind the reader that $t^2$ terms are included as in eqs. \ref{fzeroa} and \ref{fplus}.
We note that using eq. \ref{fzeroa}, suggested in \cite{stern},
the value of the phase space integral changes by only 0.04\%.
We find $I(K_{\mu 3})=0.1026 \pm 0.0005$.

 Finally, from the Callan-Treiman relation we compute $f(0) = 0.964\pm0.023$ using 
 $f_K/f_\pi = 1.189\pm 0.007$ from Ref.\cite{MILC}. Our value for  $f(0)$
 is in agreement with recent lattice calculations\cite{f0:lattice}.

\section{Conclusions}
 A new measurement of the  \kl\ \to\ $\pi\mu\nu$ form factors has been
 performed. Our  result ${\kl}_{\ell3}$ is in agreement with 
 recent measurement from KTeV\cite{KTeV:ff} and ISTRA+\cite{istra2} and in disagreement with NA48
\cite{NA48:ff2,NA48:ff}. We also derive $f(0) = 0.964\pm0.023$  in agreement with
recent lattice calculations\cite{f0:lattice}. This agreement reinforces the credibility of the $f(0)$ and $\lambda_0$ determinations.
 
\appendix
\section{Error estimates}
It is quite easy to estimate the ideal error in the estimation of a set of parameters {\bf p}=$(p_1,\:p_2,\ldots\:p_n)$ from fitting some distribution function to experimentally determined spectrum.
Let $F(\mathbf{p},x)$ be a probability density function, PDF, where ${\mathbf p}$ is some parameter vector, which we want to determine and $x$ is a running variable, like $t$. The inverse of the covariance matrix for the maximum likelihood estimate of the parameters is given by \cite{hcra}:
$$({\bf G}^{-1})_{ij}=-{\partial^2 \ln L\over\partial p_i\partial p_j}$$
from which, for $N$ events, it trivially follows:
$$\left({\bf G}^{-1}\right)_{ij}=N\:\int{1\over F}\,{\partial F\over\partial p_i }\,{\partial F\over\partial p_j}\:\dif\upsilon,$$
with $\dif\upsilon$ the appropriate volume element.
We use in the following the above relation to estimate the errors on the form factor parameters for one and two parameters expression of the form factors $\tilde f_+(t)$ and $\tilde f_0(t)$. While the errors in any realistic experiment will be larger than our estimates, typically two to three times larger, it is still very important toward the understanding of the real problems in the determination of the parameters in question.
\subsection{$K_{e3}$ decays}
For a quadratic FF, $\tilde f(t)=1+\lamp(t/m^2)+(\lampp/2)(t/m^2)^2$, the inverse of the covariance matrix ${\bf G}_+^{-1}$, the covariance matrix ${\bf G}_+$ and the correlation matrix are:
$$N\,\pmatrix{5.937 & 13.867\cr\noalign{\vglue-4mm}\cr 13.867 & 36.2405\cr},\kern5mm
{1\over N}\,\pmatrix{1.258^2 & -0.606 \cr\noalign{\vglue-4mm}\cr -0.606 & 0.509^2},\kern5mm
\pmatrix{1 & -.945\cr\noalign{\vglue-4mm}\cr  & 1\cr}$$
The square root of the diagonal elements of ${\bf G}_+$ gives the errors, which for one million events are \dlp=0.00126, \dlpp=0.00051. The correlation is very close to \minus1, meaning a fit can trade \lamp\ for \lampp\ and that the errors are enlarged.  
A fit for a linear FF, $\tilde f(t)=1+\lamp(t/m^2)$ in fact gives \lamp=0.029 instead of 0.025 and an error smaller by \ab3:
$$\dlp=\sqrt{\left({1\over N}\:G_+^{-1}(1,1)\right)^{-1}}=0.0004.$$
A simple rule of thumb is that ignoring a $t^2$ term, increases \lamp\ by \ab3.5\x\lampp.
For $K_{e3}$ decays the presence of a $t^2$ term in the FF is firmly established. It is however incorrect to try to fit for two terms connected by the simple relation \lampp=2\x\lamp\up2, both from theory and experiment. The above discussion justifies the use of eq. \ref{fplus}. The errors obtained above compare reasonably with the errors quoted in \cite{KTeV:ff,NA48:ff,KLOE:ff}.
\subsection{$K_{\mu3}$ decays}
The scalar FF only contributes to $K_{\mu3}$ decays. Dealing with these decays is much harder because: a) - the branching ratio is smaller, resulting in reduced statistics, b) - the $E_\pi$ or $t$ range in the decay is smaller, c) - it is in general harder to obtain an undistorted spectrum and d) - more parameters are necessary. This is quite well evidenced by the wide range of answers obtained by different experiments \cite{KTeV:ff,istra2,NA48:ff2}.
Assuming that both scalar and vector FF are given by quadratic polynomials as in eq. \ref{eq:ff2}, ordering the parameters as \lamop, \lamopp, \lamp\ and \lampp, the matrices ${\bf G}_{0\,\&\,+}^{-1}$ and ${\bf G}_{0\,\&\,+}$, are:
$$N\pmatrix{1.64&5.44&1.01&3.90\cr\noalign{\vglue-4mm}\cr
5.44&18.2&3.01&12.3\cr\noalign{\vglue-4mm}\cr
1.01&3.01&1.47&4.24\cr\noalign{\vglue-4mm}\cr
3.90&12.3&4.24&13.8\cr},\ {1\over N}\pmatrix{63.9^2&-1200&-923&197\cr\noalign{\vglue-4mm}\cr
-1200&18.8^2&272&-59\cr\noalign{\vglue-4mm}\cr
-923&272&14.8^2&-49\cr\noalign{\vglue-4mm}\cr
197&-59&-48&3.4^2 \cr}$$
and the correlations, ignoring the obvious terms, are:
\begin{equation}
\label{4corr}
\pmatrix{-0.9996&-0.974&0.91\cr\noalign{\vglue-4mm}\cr
 &0.978&-0.919\cr\noalign{\vglue-4mm}\cr
 & &-0.976\cr}.
\end{equation}
All correlations are very close to \minus1. In particular the correlations between \lamop\ and \lamopp\ is \minus99.96\%, reflecting in vary large \dop\ and \dopp\ errors. We might ask what the error on \lamop\ and \lamopp\ might be if we had perfect knowledge of \lamp\ and \lampp. The inverse covariance matrix is give by the elements (1,1), (1,2), (2,1) and (2,2) of the ${\bf G}_{0\,\&\,+}^{-1}$ matrix above. The covariance matrix therefore is :
$${\bf G}_{0}(\lamop,\ \lamopp\ \hbox{for \lamp, \lampp\ known})={1\over N}\pmatrix{8.2^2&-20\cr\cr\noalign{\vglue -4mm}-20&2.4^2\cr}.$$
For one million events we have \dopp=0.0024, about 4\x\ the expected value of \lamopp. In other words \lamopp\ is likely to be never measurable. It is however a mistake to use a scalar form factor linear in $t$, because the coefficient of $t$ will absorb the coefficient of a $t^2$ term, again multiplied by \ab3.5. Thus a real value \lamop=0.014 is shifted by the fit to 0.0161, having used eq. \ref{fzeroa}. Fitting the pion spectrum from 1 million $K_{\mu3}$ decays for \lamo, \lamp\ with the form factors of eq. \ref{fzeroa} gives the errors \dop\ab0.001 and \dlp\ab0.0011. Combining with the result from a fit to 1 million $K_{e3}$ with the FF of eq. \ref{fplus} for which \dlp\ab0.0004 gives finally \dop\ab0.00082, \dlp\ab0.00038 and a \lamo-\lamp\ correlation of \minus29\%. We hope to reach this accuracy with our entire data sample, \ab5\x\ the present one, and a better analysis which would allow using the pion spectrum.
\section*{Acknowledgements}
We thank the DA$\Phi$NE team for their efforts in maintaining low background running 
conditions and their collaboration during all data-taking. We want to thank our technical staff: 
G.F.Fortugno for his dedicated work to ensure an efficient operation of the KLOE Computing Center; 
M.Anelli for his continous support to the gas system and the safety of the detector; 
A.Balla, M.Gatta, G.Corradi and G.Papalino for the maintenance of the electronics;
M.Santoni, G.Paoluzzi and R.Rosellini for the general support to the detector; 
C.Piscitelli for his help during major maintenance periods.
This work was supported in part by DOE grant DE-FG-02-97ER41027; 
by EURODAPHNE, contract FMRX-CT98-0169; 
by the German Federal Ministry of Education and Research (BMBF) contract 06-KA-957; 
by Graduiertenkolleg `H.E. Phys. and Part. Astrophys.' of Deutsche Forschungsgemeinschaft,
Contract No. GK 742; by INTAS, contracts 96-624, 99-37; 
by TARI, contract HPRI-CT-1999-00088.

\end{document}